# Fit of Fossils and Mammalian Molecular Trees: Dating Inconsistencies Revisited

**Peter J. Waddell**[1]

pwaddell@purdue.edu

[1] Department of Biological Sciences, Purdue University, West Lafayette, IN 47906, U.S.A.

Divergence time estimation requires the reconciliation of two major sources of data. These are fossil and/or biogeographic evidence that give estimates of the absolute age of nodes (ancestors) and molecular estimates that give us estimates of the relative ages of nodes in a molecular evolutionary tree. Both forms of data are often best characterized as yielding continuous probability distributions on nodes. Here, the distributions modeling older fossil calibrations within the tree of placental (eutherian) mammals are reconsidered. In particular the Horse/Rhino, Human/Tarsier, Whale/ Hippo, Rabbit/Pika and Rodentia calibrations are reexamined and adjusted. Inferring the relative ages of nodes in a phylogeny also requires the assumption of a model of evolutionary rate change across the tree. Here nine models of evolutionary rate change, are combined with various continuous distributions modeling fossil calibrations. Fit of model is measured both relative to a normalized fit, which assumes that all models fit well in the absence of multiple fossil calibrations, and also by the linearity of their residuals. The normalized fit used attempts to track twice the log likelihood difference from the best expected model. The results suggest there is a very large difference in the age of the root proposed by calibrations in Supraprimates (informally Euarchontoglires) versus Laurasiatheria. Combining both sets of calibrations results in the penalty function vastly increasing in all cases. These issues remain irrespective of the model used or whether the newer calibrations are used.

## 1   Introduction

This research follows as a rejoinder to a series of earlier work. In particular, the reader is referred to Waddell (2007) and references therein. In particular, the question of why fossil calibrations within the superorder Supraprimates (Waddell et al. 2001, informally called 'Euarchontoglires' after Waddell et al. 1999c) clash with those within the superorder Laurasiatheria (Waddell et al. 1999c) is considered in more detail. This includes reexamining and updating the best of the fossil calibrations themselves. It also involves applying a set of nine different models and some newly applied measures of fit of data to model to measure how well fossil and molecular data fit together.

By fossil calibration is meant the combined sum of knowledge based on geological sciences, with minimal direct inference from other sources of data, which allows a reasonable inference of the age of a lineage(s) (Waddell and Penny 1996). The usable form of this inference is a probability distribution on the age of a node. This will typical be some form of continuous distribution, reflecting the knowledge accumulated by expert paleontologists regarding the origin of particular extant lineages. This should include a great deal of relevant expertise, such as relevant sites that do not yield fossils, which may not make it into the literature, but which can be evaluated by personal correspondence. Thus, a fossil calibrated node is a node in the molecular tree for which there is reasonable chronological data. The older usage of fossil calibration point is ambiguous; it may mean a fossil calibrated node, or more particularly, the practice of treating the fossil calibration distribution as a single point or delta function. Unfortunately, no historical inference can be that accurate.

## 2  Materials and Methods

The weighted tree of placental mammals, shown again in appendix 1, is the same as that used in Waddell (2007). In terms of data, methods and many of the questions addressed, this research may be read as an extension of Waddell (2007) so readers are referred there for further details. Revised modeling within this work, for example, inclusion of a log normal or exponential distribution of inferred fossil calibrations, used Excel (2003, 2004), along with the included data analysis packages including Solver (Frontline systems, 2004).

## 3  Results
### 3.1 How fossil calibrations are derived

For divergence time studies, only part of the fossil record is usually directly considered. This information is used to estimate a distribution of the probability that a node in the molecular tree is a particular age (Waddell and Penny 1996). This typically requires fossils of known age that appear, due to derived features (synapomorphies), to place a strong lower limit on the age of a lineage in the tree. This is effectively the same as saying they place a minimum age on a node in the tree. Ideally, there will be a fairly well sampled and continuous series of fossils near the node. That is, it is expected with high probability that the ancestors (or individuals very close to them) have been sampled, and that this sampling is fairly fine grained (not too many gaps and discontinuities), that both descendant lineages are also sampled, and there are wide ranging fossil sites (in space and time) at which the taxa of interest should be found. There also needs to be consideration of the difficulty of identifying the earliest members of a lineage that may not show clear synapomorphies, and the converse issue of mistaking a member of an ancestral lineage for something more derived. At present, these alternative concerns need to be balanced, using subjective, but hopefully, expert opinions. In the best situations, the fossil record alone can lead to a fairly tight and symmetric distribution for the node time. An example of this is the normally distributed estimate of the age of the horse/rhino (+ tapir) split at 55 mya with standard deviation of 1.5 myr proposed by Waddell et al. (1999a) subsequently widely adopted in other studies without major modification. Another is the age of the African ape/Orangutan split. Waddell and Penny (1996) assumed and used a calibration of 16 million years before present (mybp) with s.d. of 1 million years (myr) based on their interpretation of published data. This was somewhat older than most contemporary estimates that ranged from about 10 to 14 mybp, however, it has been adopted by other molecular studies. It is corroborated by more recent research (e.g., Begun 2003).

The form of the distribution for a nodes age may well be conceived as non-symmetric (e.g., Waddell et al. 2001). If, for example, only a few scattered fossils showing what are interpreted as clear synapomorphies are found, but without a good sampling of probable ancestors, then a minimum date for a lineage may seem fairly well established, but the upper date remains very lose. A good example of within placentals is the age of the armadillo lineage. There is a fossil scute dated at about 61 million years ago (mya) from South America (McKenna and Bell 1997,and references therein). Most experts would say, based on this alone, there is a 50% or better probability that the armadillo lineage is at least this old. That is, few would seriously doubt the dating of the fossil, nor the assignment of it to the armadillo lineage. However, there is a very poor sampling of other scutes in the early Eocene/Paleocene, suggesting it is unlikely this single occurrence accurately tracks the first appearance of this feature, at least not without making further assumptions. A significant issue is that there are no sites known with probable ancestors of Xenathra. Taking this evidence by itself, suggests the inferred age of the Xenathra node should be pretty much unbounded above. Hence, the distribution may be viewed as highly asymmetric, some might argue a translated exponential with a fairly large standard deviation. Coincidently, under some strict assumptions, which are generally not met with mammal fossils, the distribution is exactly this (Marshall 1990).

The previous example also raises some interesting questions. Is the distribution for the age of Xenathra bounded above? Some would say yes, it logically cannot be older than the root of the placentals (of which it is a member). While there is no direct fossil evidence for the age of placentals, everyone has expectations of when it might have been. The point to make here is there needs to be considerable caution with regards to letting such nebulous expectations directly into fossil calibrations.

There are, however times when fossils not showing derived synapomorphies do seem to have a clear bearing on the expected age of a node under consideration. A good example of this is again the horse/rhino (+ tapir divergence). Morphological and molecular evidence both strongly support rhino + tapir as more closely related to each other than to horse (e.g., Waddell and Shelly 2003). The fossil record was good enough in this case that it seemed to track this divergence in North America. That is, at about 55 mybp with a standard deviation of about 1.5myr and a roughly normal distribution, horse split from the rhino/tapir ancestor. About 3myr later (52 mya +/- 1.5myr) there is clear evidence of tapir splitting from rhino (e.g., McKenna and Bell 1997, and references therein). It is tempting to treat these as independent calibrations, but they clearly are not. It is logically impossible for the later date to precede the former, so there would seem to be a truncated bivariate distribution. Also, there is an expectation that the record suggests about 3 myr between the two events, during which rhino/tapir ancestors seem to be identified. This in turn suggests the joint distribution may be modeled as a truncated bivariate normal with some positive correlation. The change in twice the log likelihood of such joint normal data can be measured using a generalized sum of squares, that is $ss = \mathbf{dV^{-1}d}^t$, where d is a vector of the differences of the proposed divergence time from its expected values and $\mathbf{V}$ is the variance-covariance matrix of the expected ages. This is not used in the analyses below, because the molecular tree used does not include the tapir lineage, but is likely to become a more common issue in densely sampled trees.

Something that is truly undesirable is contamination of fossil / paleontological / biogeographic data with molecular data. Morphological cladistics can be a notoriously subjective exercise. For example, despite all the confidence demonstrated in high profile publications such as Novacek (1992), the tree relating placental orders present was nearly completely wrong. It has been replaced with the tree of Waddell et al. (1999c), a set of hypotheses now strongly corroborated by multiple studies. A danger is that in the aftermath of novel molecular phylogenetic evidence, paleontologists may go in the opposite direction and seek to reinterpret the morphological data and paleontological data to fit molecular expectations. It can thus become very unclear, in the absence of any objective model, what the morphological/paleontological data are independent of molecular sensibilities. A similar thing can happen with divergence time estimates. Molecular estimates may be published in popular articles that are faulty in their own ways, but are uncritically widely reported and followed. Paleontologists that have poor critical understanding of molecular phylogenetics, pick up on such articles and will either treat them as dogma, ignore them or strongly dispute them. In so doing they bias their own interpretation of the fossils.

**3.2 Revised interpretations of deep placental calibration points**

A range of fossil calibrations have been used to infer the age of the root of placental mammals using molecular trees. A majority of these come from Waddell (1999a) and Waddell et al. (2001). These have been adopted fairly widely (e.g., Springer et al. 2003, c.f. Kitazoe et al. 2007). The way that these calibrations were originally devised was from a combination of my own reading of the paleontological literature, typically followed up with focused and sometimes lengthily discussions/interviews with prominent fossil experts. These include widely published paleontologists such as Chris Beard, Mary Dawson, David Archibald and Hans Thewissen. To clarify, I do not take their opinions directly, but I certainly mine them for information on aspects of the fossil record that are important, but poorly reported. These include how many fossil stratigraphies, localities and regions that could be expected to include the taxa of interest have been sampled adequately, hand how well the ancestors seem to be known. As noted elsewhere (Waddell et al. 2001) these distributions represent expert opinion. Like all Bayesian priors they presently include a considerable element of subjectivity, but as yet there is not a better method and they cannot be circumvented.

Revisions that have some merit for being older than they currently are include:

Horse/Rhino or HR (+ tapir). This calibration is first developed in Waddell et al. (1999a). There the Horse/Rhino split is treated as normally distributed with mean 55 mybp and s.d. 1.5 myr (that is $\mu$ = 55, $\sigma^2$ = 1.5$^2$). The rhino tapir split has a mean of 52 mybp and s.d. 1.5 myr. There is good fossil data from North America before and after these events. There has however in the last 10 years been evidence to suggest that animals virtually indistinguishable from the ancestors were more widespread and appeared earlier elsewhere and also that some to the derived features were appearing amongst these fossils earlier than previously appreciated (Chris Beard pers. comm.). One of the important points is that the perissodactyl lineage appears to be migrating from Asia to North America at this time (Beard 1998). Given this and other recent findings, then the expected age of the horse rhino split might be better placed at 58 mya, and that of rhino/tapir at about 55 mya (Chris Beard pers. comm., based on new data from Asia in the past decade). The variances of these times adopted in Waddell (1999a) are still used, but may need to be revised. Anticipating the horse / rhino / tapir divergence this as a bivariate calibration in future, the correlation of these two times is assumed to be about 0.7, while mean time for the tapir split is revised upwards to 55 mybp.

Whale/Hippo or WH: This time was inferred to be in the range 49 to 61 mya with 95% confidence by Waddell et al. (2001) with a mean of 52 myr. It is preferable in many ways to estimate the median or the modal time, since the mean is much harder to subjectively judge if a long tailed distribution is to be fitted. If we will equate the previous estimate of the mean with the median, this comes close to a translated lognormal distribution with parameters $\sigma^2$ = 0.8893, $\mu$ = 0.7729, y = 48.62. The fitting was done using an Excel spread sheet, invoking the Solver function (Frontline systems, 2004) to maximize the fit to the prescribed points of the distribution.

More recently discovered fossils show derived whale-like features appearing somewhat earlier than previously expected (e.g., Thewissen et al. 2001, Nummela et al. 2006). There also seems to be a mode, or most probable time, for this event near to 51 or 52 mybp, assuming the transition to water from the common hippo-whale ancestor occurred fairly rapidly as seems to be indicated by the fossils. Fitting a log normal to have a mode of 51.47 mybp and 2.5% and 97.5% points of 50 and 65 mybp respectively, yields a translated lognormal with parameters $\sigma^2$ = 0.51613, $\mu$ = 0. 1.8383, y = 47.71. The upper point is very hard to judge, however there are earlier sites which might be expected to show whale-like fossils or potential ancestors, which do not (Hans Thewissen, pers. comm.). Fitting continuous distributions such as a log-normal involves compromises. In this case, to fit the required mode and lower 2.5% point, and to have a reasonable absolute lower bound (the y value), the 97.5% point needed to made older than previously expected. This results in a log normal with perhaps a longer thicker tail than might be desirable. However, it matters relatively little in these analyses, since the whale calibration invariably always pushed towards its absolute lower limit, which should be close to 48 mybp given abundant fossils of that age or older that can only realistically be called whale ancestors.

Rabbit/Pika or RP: Following Waddell et al. (2001), this was estimated to have a lower 2.5% point of 36, and upper 97.5% point of 55 and a mean of 42.5. This was fitted with a long normal distribution with parameters $\sigma^2$ = 0.6323, $\mu$ = 1.7928, y = 34.26. Since the mode of this distribution is 37.45, it was refitted so that the mode was 41 mybp, as this seems more intuitively reasonable. This required the upper 97.5% point to be shifted to 62 mybp to fit perfectly. This second fitting yielded the parameters $\sigma^2$ = 0.1502, $\mu$ = 3.7733, y = 3.5763, a less skewed distribution. As yet there is no new paleontological data to strongly narrow the wide confidence interval on this calibration (Mary Dawson and Chris Beard, pers comm.).

Rodentia: In Waddell (2001) this calibration was considered but not used. There is now a lot more paleontological data relevant to this question. A wide variety of Asian sites are now being studied (Dawson 2003, Wible et al. 2005, Meng et al. 2007). Given a wide variety and quick succession of early Glires forms it appears most likely that Rodentia evolved in Asia and moved into North America soon thereafter. The general view is that Asian sites are now yielding a fairly close succession of fossils close to the origin of rodents (Mary Dawson and Chris Beard, pers. comm.). There appears to be enough data to make an estimate like that made earlier for Horse / Rhino. In this case for Rodentia it would be 60 mybp with a s.d. of 1.5 myr. This could be further modified with a lower bound of ~57mybp. However, it is the upper bound that appears much more valuable. Rodent teeth are highly diagnostic of the crown group Rodentia, and are clearly differentiated from the ancestral teeth of Glires. Their first occurrence puts any upper bound on the age of Rodentia. At present this is about 57mybp. Despite many fossil sites scattered across Asia and North America, and despite rodents seeming to become widespread soon after their first occurrence, these teeth have not been found at any of many older sites. If there is a deep calibration within Placentalia that appears to come close to Marshall's (1990) model, then this is it. Indeed, future work could focus on quantifying data for this model. At present, the age of Rodentia might also be modeled as a translated exponential with a lower bound (translation) of 60 mybp and a standard deviation of 1.5 myr. This is constructed to favor older times for this node, and if anything, could be shifted back a couple of million years (e.g., translation factor 58 mybp).

Human/Tarsier or HT: The direct fossil estimates of the first tarsiers continue to point to close to 55 mybp (Chris Beard per comm., Beard 1991, 2008, Beard et al. 2007, Rossie et al. 2006). Based on paleontological data alone, this calibration remains reasonable. However, it does seem to perturb the fit on the tree in the analyses below, therefore, when assuming older times for fossil calibrations, a mean of 60 mybp is used instead of 55 mybp (Waddell et al. 2001). As seen in various analyses, such a calibration is perfectly compatible with the ages of Rodentia and Rabbit / Pika described above. To again be very clear, this older calibration is being used for purely exploratory purposes.

**3.3 The fitting process**

For the assessments below the 2-way penalty described in Waddell (2007) was used for the random walk models of rate change, namely R2, L2 and I2, which are respectively, a random walk on the evolutionary rate, the log of the rate and the inverse of the rate. For the Brownian models, namely RT, LT, IT, RE, LE and IE, where T indicates that the variance of the process is proportional to time, while E indicates it is proportional to the product of rate and time (the edge length), a 3-way penalty was used. All models used the scale factor, which corrects for the unknown variance of the stochastic process, as described in Waddell (2007) table 3. All penalties across the root used the appropriate weighted average described in Waddell and Kalakota (2007). For each method the best fit of the tree in Waddell (2007) was measured with the Horse/Rhino calibration set to 55 mybp (due to the scale factor, it makes no difference if 58mybp is used instead). The fit was then scaled to the degrees of freedom of the system (in this case, 32). This follows the logic of Kitazoe et al. (2007), which assumes that the data fit the model, in order to avoid the need to estimate the variance of the stochastic process. This normalized fit may be regarded as proportional to minus twice the log likelihood of the model.

The penalty for the molecular node time moving away from the mode of the fossil calibration distributions was also estimated as twice the decrease in the log likelihood. For the normal distribution, this is equal to $\frac{(x-\mu)^2}{\sigma^2}$. For the log normal distribution, the term is $2\ln\left[\frac{1}{x}\right] - \frac{(\ln[x]-\mu)^2}{\sigma^2}$ minus the value of this function at the mode of the distribution (which occurs at $e^{\mu-\sigma^2}$ and was calculated numerically). For the exponential it is $2\left(\ln\left[\frac{1}{\lambda}e^{-\lambda x}\right]\right)$ minus the value at the mode of this function, that is when x = 0. For the last two distributions, x is evaluated after any translation (y) has already been subtracted. If the fossil calibrations coincide exactly with the relative values estimate by the molecular model, then the additional penalty over the molecular penalty (which is normalized to 32) will be zero.

The fit of data to model is evaluated by measuring the linearity of the standardized residuals compared to their expected values (Waddell 2007). This is basically the linearity of a Q-Q plot. The absolute quality of this fit was assessed using the tables in Filliben (1975). In particular, the test of Filliben sorts the observed residuals against the order statistic medians and then measures the Pearson Correlation between them. The order statistic medians are obtained as the value of the normal distribution for which exactly half of that order statistic are expected to be larger and half smaller in value. In Excel, for example, it is obtained with the function =NORMINV((x-0.5)/n,0,1), where x is the order statistic of interest, n is the total number of observations, while 0 is the mean and 1 the standard deviation of a standardized normal distribution.

**3.4 Inferences using Supraprimates calibrations only**

Each of the nine models were fitted using the older set of calibration times within Supraprimates. These are the Human Tarsier (normal, $\mu$ = 60, $\sigma^2$ = 2.5$^2$), Rabbit Pika (log normal, $\sigma^2$ = 0.1502, $\mu$ = 3.7733, y = 3.5763) and Rodentia (translated exponential, translation y = 60 myr, $\lambda$ = 1.5myr) calibrations. The results are shown in table 1.

Table 1 shows some interesting features. For all models, the additional stress of adding multiple calibrations was minimal. While the HT calibration contributed most to the stress on the fossil times, this is only because the Rodentia age was bounded below by 60 mybp. If the normal calibration described above was used for Rodentia instead, the fit became even better and the age of the HT and Rodentia nodes became ~1 myr less (results not shown). The use of the exponential distribution for Rodentia introduce optimization problems, such that, to arrive at a good fit, it had to be reset to above its boundary value (e.g., to 61 mybp) when a new optimization was begun. Use of the normal distribution here did not have this problem, but the results (not shown) were very similar.

It is interesting that all models infer the root of placentals to be ~82-85 mybp, which is quite compatible with what most paleontologists would expect. This compatibility is even more so when it is realized all of these models allow a confidence interval of ~ +/- 5myr due to all the variables presented here (calibration uncertainty plus variance of the rate of evolution model). As noted previously (Waddell 2001, Waddell 2007), the fit with calibrations in Laurasiatheria is very poor for all models. The Cetacea calibration is not actively used in this article, but assumes that the split of sperm from baleen whales occurred 34 mybp with s.d. 4 myr. The predictions of ages within Laurasiatheria is best using the RT model, but remains very poor. Fossil ages within Afrotheria are relatively poorly known, and the edges in this molecular tree do not form a tight bound, as it lacks sirenian sequences, for example. None of the ages here break assumed calibrations, which are that the elephant lineage is probably older than 55 mybp while the Xenathra lineage is probably greater than 61 mybp. Within Laurasiatheria, not only the calibration nodes, but also some of the deeper groups, specifically chiropteran (bat) ages seem somewhat too young, although fossil data near the root of Chiroptera remains controversial. Ages for Eulipotyphla seem reasonable given the strong skepticism for any fossils definitely in this group dating from prior to the Paleocene.

Table 2 shows the fit of the fit of the residuals of these models to expected values and each other. As seen in Waddell (2007) the LE and IT models fit very well. Tables in Filliben (1975) suggest that a correlation of greater than 0.97 indicates the expected fit. However, it may be desirable to confirm this with simulations given the numerical optimization. It is also possible that this is a rather weak test of fit of data to model.

Table 1. Fitting multiple models of the rate of evolution of the rate of evolution to the tree of placental mammals. The HT, RP and Rodentia nodes had the older fossil calibrations. The fit (twice ln L difference from optimal fit) of other fossil calibrations is also shown, but these were not part of the penalty function here (-1000 indicates the WH is below its minimum allowed value). Model fit is relative to a single calibration normalized to 32. Ages in mybp. Underlined are nodes with well described calibrations. The fit to the Cetacea calibration is shown, but never used during the optimizations within this article.

| Calibration\Model | R | L | I | RE | LE | IE | RT | LT | IT |
|---|---|---|---|---|---|---|---|---|---|
| HR | -189.0 | -206.8 | -229.6 | -127.5 | -169.7 | -201.4 | -101.2 | -139.7 | -181.9 |
| HT | -0.1 | -0.2 | -0.2 | -0.3 | -0.3 | -0.4 | -0.2 | -0.3 | -0.3 |
| WH | -1000 | -1000 | -1000 | -1000 | -1000 | -1000 | -1000 | -1000 | -1000 |
| RP | -0.1 | -0.1 | -0.1 | -0.1 | -0.1 | -0.1 | -0.1 | -0.1 | -0.1 |
| Rodentia | 0.0 | 0.0 | 0.0 | 0.0 | 0.0 | 0.0 | 0.0 | 0.0 | 0.0 |
| Cetacea | -22.5 | -26.1 | -29.6 | -22.2 | -29.2 | -34.8 | -19.3 | -25.2 | -32.3 |
| Model | R | L | I | RE | LE | IE | RT | LT | IT |
| Model Fit | 32.37 | 32.49 | 32.62 | 32.65 | 32.77 | 32.88 | 32.57 | 32.62 | 32.73 |
| Clade ages | R | L | I | RE | LE | IE | RT | LT | IT |
| Placentalia/Root | 82.04 | 82.82 | 83.43 | 82.97 | 84.16 | 85.08 | 81.83 | 82.78 | 84.18 |
| Boreotheria | 76.88 | 77.35 | 77.68 | 77.48 | 78.26 | 78.84 | 76.67 | 77.30 | 78.23 |
| Atlantogenata | 78.85 | 79.48 | 79.97 | 79.76 | 80.75 | 81.50 | 78.83 | 79.62 | 80.76 |
| Laurasiatheria | 68.86 | 68.55 | 68.05 | 68.40 | 68.29 | 68.10 | 68.24 | 68.16 | 68.10 |
| Supraprimates | 71.08 | 71.41 | 71.66 | 71.54 | 72.00 | 72.33 | 71.04 | 71.41 | 71.95 |
| Afrotheria | 64.17 | 63.96 | 63.59 | 63.87 | 63.80 | 63.66 | 63.85 | 63.99 | 63.80 |
| Scrotifera | 64.69 | 64.15 | 63.37 | 64.09 | 63.55 | 63.02 | 64.23 | 63.81 | 63.30 |
| Eulipotyphla | 60.73 | 59.49 | 58.00 | 58.00 | 56.66 | 55.53 | 58.80 | 57.77 | 56.47 |
| Euarchonta | 69.75 | 70.06 | 70.29 | 70.23 | 70.64 | 70.93 | 69.79 | 70.11 | 70.59 |
| Glires | 66.25 | 66.48 | 66.67 | 66.55 | 66.80 | 66.99 | 66.25 | 66.46 | 66.75 |
| Afroinsectiphilla | 60.07 | 59.52 | 58.80 | 59.26 | 58.86 | 58.45 | 59.60 | 59.51 | 58.97 |
| Fereuungulata | 61.52 | 60.87 | 59.96 | 60.99 | 60.19 | 59.46 | 61.34 | 60.67 | 59.89 |
| Chiroptera | 51.19 | 50.32 | 49.06 | 50.08 | 48.53 | 46.99 | 50.91 | 49.68 | 48.03 |
| Primates | 62.23 | 62.41 | 62.56 | 62.60 | 62.75 | 62.89 | 62.48 | 62.56 | 62.73 |
| Lagomorpha | 44.54 | 44.69 | 44.80 | 45.41 | 45.29 | 45.21 | 45.42 | 45.60 | 45.25 |
| Rodentia | 60.00 | 60.00 | 60.00 | 60.00 | 60.00 | 60.00 | 60.00 | 60.00 | 60.00 |
| Ferae | 57.30 | 56.60 | 55.63 | 56.73 | 55.63 | 54.69 | 57.31 | 56.37 | 55.30 |
| Euungulata | 58.19 | 57.39 | 56.39 | 57.98 | 56.92 | 56.01 | 58.52 | 57.60 | 56.58 |
| megabats | 23.36 | 22.29 | 21.13 | 23.59 | 21.55 | 19.72 | 24.87 | 22.83 | 20.78 |
| human/tarsier | 60.55 | 60.67 | 60.77 | 60.83 | 60.92 | 61.01 | 60.78 | 60.80 | 60.90 |
| hystrich./murid_ | 52.89 | 52.72 | 52.55 | 51.44 | 51.00 | 50.60 | 51.90 | 51.56 | 51.12 |
| Carnivora | 41.52 | 40.97 | 40.10 | 41.88 | 40.58 | 39.51 | 42.76 | 41.59 | 40.30 |
| Perissodactyla | 41.16 | 40.39 | 39.44 | 44.17 | 42.04 | 40.62 | 45.68 | 43.52 | 41.48 |
| Cetartiodactyla | 45.34 | 44.11 | 42.99 | 45.52 | 43.29 | 41.81 | 46.78 | 44.66 | 42.66 |
| Anthropoidea | 36.69 | 36.09 | 35.46 | 35.08 | 33.76 | 32.45 | 35.78 | 34.57 | 33.21 |
| mouse/rat | 14.93 | 14.80 | 14.56 | 14.13 | 13.72 | 12.74 | 14.33 | 13.96 | 13.42 |
| Caniformia | 33.93 | 33.49 | 32.75 | 34.23 | 33.06 | 32.10 | 35.05 | 34.01 | 32.83 |
| Artiofabula | 41.47 | 40.20 | 39.08 | 41.70 | 39.32 | 37.77 | 43.07 | 40.77 | 38.66 |
| Cetruminantia | 35.64 | 34.32 | 33.22 | 35.80 | 33.28 | 31.68 | 37.25 | 34.79 | 32.57 |
| Whippomorpha | 31.77 | 30.42 | 29.33 | 31.91 | 29.36 | 27.74 | 33.38 | 30.88 | 28.62 |
| Cetacea | 15.02 | 13.55 | 12.23 | 15.13 | 12.40 | 10.39 | 16.41 | 13.90 | 11.25 |

Table 2. Correlation of residuals of the nine models calibrated with the Supraprimates fossils to their expected median values and to each other. Highlighted are the two highest fits with respect to expected residuals.

|      | Expt. | R | L | I | RE | LE | IE | RT | LT |
|------|-------|-------|-------|-------|-------|-------|-------|-------|-------|
| Expt. | 1.000 | | | | | | | | |
| R    | 0.946 | 1.000 | | | | | | | |
| L    | 0.963 | 0.992 | 1.000 | | | | | | |
| I    | 0.976 | 0.984 | 0.994 | 1.000 | | | | | |
| RE   | 0.975 | 0.978 | 0.988 | 0.990 | 1.000 | | | | |
| LE   | 0.982 | 0.966 | 0.981 | 0.989 | 0.993 | 1.000 | | | |
| IE   | 0.988 | 0.953 | 0.973 | 0.984 | 0.984 | 0.990 | 1.000 | | |
| RT   | 0.973 | 0.981 | 0.987 | 0.984 | 0.996 | 0.985 | 0.979 | 1.000 | |
| LT   | 0.978 | 0.971 | 0.987 | 0.992 | 0.996 | 0.995 | 0.987 | 0.988 | 1.000 |
| IT   | 0.986 | 0.968 | 0.983 | 0.989 | 0.983 | 0.983 | 0.992 | 0.981 | 0.983 |

## 3.5 Inferences using Laurasiatheria calibrations only

The analyses in section 3.4 were repeated, using only the older WH and HR calibrations. Table 3 shows that most of the models show a substantially worse fit, indicating that many of them detect an incompatibility of these two fossil calibrations with respect to the molecular tree. The root appears to be uniformly old at ~125 to 129 mybp. Table 4 shows that the models with the better residuals do not show the best fit statistics. This may be a symptom of assuming the fit of all models is equivalent when normalizing the residual penalty to 32.

Table 3. Fitting multiple models of the rate of evolution of the rate of evolution to the tree of placental mammals. The HR and WH nodes used the older fossil calibrations. The fit (twice ln L difference from optimal fit) of other fossil calibrations is also shown, but these were not part of the penalty function. Model fit is relative to a single calibration normalized to 32. Ages in mybp. Underlined are nodes with well described calibrations.

| Calibration\Model | R | L | I | RE | LE | IE | RT | LT | IT |
|------|------|------|------|------|------|------|------|------|------|
| HR | 0.0 | -0.2 | -0.6 | -0.1 | -0.8 | -1.7 | 0.0 | -0.3 | -1.1 |
| HT | -493.9 | -454.9 | -451.5 | -498.5 | -457.1 | -430.9 | -457.9 | -463.9 | -414.7 |
| WH | -0.1 | -0.5 | -1.0 | -0.3 | -1.2 | -1.8 | -0.1 | -0.6 | -1.4 |
| RP | -2.1 | -2.0 | -2.0 | -2.2 | -2.0 | -1.9 | -2.1 | -2.1 | -1.9 |
| Rodentia | -482.6 | -434.2 | -423.5 | -472.8 | -425.6 | -396.0 | -439.1 | -441.5 | -385.5 |
| Cetacea | -6.1 | -8.3 | -10.5 | -6.3 | -9.8 | -13.4 | -5.0 | -7.8 | -11.9 |
| Model | R | L | I | RE | LE | IE | RT | LT | IT |
| Fit | 33.00 | 35.38 | 37.86 | 35.96 | 41.82 | 46.96 | 33.82 | 37.96 | 43.84 |
| Clade ages | R | L | I | RE | LE | IE | RT | LT | IT |
| Placentalia/Root | 128.13 | 127.19 | 127.77 | 129.27 | 128.89 | 128.68 | 125.65 | 127.37 | 126.78 |
| Boreotheria | 120.04 | 118.76 | 118.88 | 120.65 | 119.78 | 119.17 | 117.68 | 118.88 | 117.75 |
| Atlantogenata | 123.14 | 122.06 | 122.45 | 124.25 | 123.64 | 123.26 | 121.01 | 122.48 | 121.60 |
| Laurasiatheria | 107.52 | 105.22 | 103.98 | 106.39 | 104.37 | 102.74 | 104.67 | 104.74 | 102.38 |
| Supraprimates | 110.92 | 109.56 | 109.59 | 111.32 | 110.12 | 109.28 | 108.97 | 109.75 | 108.23 |
| Afrotheria | 100.21 | 98.22 | 97.40 | 99.38 | 97.64 | 96.38 | 97.89 | 98.38 | 96.05 |
| Scrotifera | 100.99 | 98.46 | 96.80 | 99.51 | 97.03 | 95.03 | 98.43 | 97.94 | 95.14 |
| Eulipotyphla | 94.82 | 91.30 | 88.58 | 89.99 | 86.43 | 83.66 | 90.08 | 88.63 | 84.81 |
| Euarchonta | 108.88 | 107.53 | 107.54 | 109.28 | 108.03 | 107.16 | 107.05 | 107.76 | 106.18 |
| Glires | 103.07 | 101.67 | 101.61 | 103.32 | 101.93 | 101.00 | 101.45 | 101.95 | 100.20 |
| Afroinsectiphilla | 93.80 | 91.40 | 90.06 | 92.18 | 90.08 | 88.51 | 91.35 | 91.49 | 88.77 |
| Fereuungulata | 96.00 | 93.41 | 91.60 | 94.47 | 91.80 | 89.67 | 93.84 | 93.00 | 90.03 |
| Chiroptera | 79.91 | 77.22 | 74.91 | 77.21 | 73.84 | 70.86 | 77.65 | 75.97 | 72.19 |
| Primates | 97.71 | 96.34 | 96.27 | 97.95 | 96.51 | 95.55 | 96.34 | 96.66 | 94.85 |
| Lagomorpha | _69.17_ | _68.19_ | _68.11_ | _69.99_ | _68.59_ | _67.74_ | _69.11_ | _69.53_ | _67.43_ |
| Rodentia | _92.95_ | _91.26_ | _90.87_ | _92.62_ | _90.95_ | _89.85_ | _91.43_ | _91.52_ | _89.45_ |
| Ferae | 89.39 | 86.86 | 84.99 | 87.51 | 84.71 | 82.53 | 87.44 | 86.22 | 83.15 |
| Euungulata | 90.64 | 88.04 | 86.21 | 89.42 | 86.67 | 84.53 | 89.28 | 88.11 | 85.08 |
| megabats | 36.46 | 34.21 | 32.26 | 36.21 | 32.73 | 29.73 | 37.80 | 34.82 | 31.23 |
| human/tarsier | _95.14_ | _93.72_ | _93.60_ | _95.30_ | _93.81_ | _92.82_ | _93.83_ | _94.05_ | _92.20_ |
| hystrich./murid_ | 81.83 | 80.01 | 79.32 | 79.05 | 76.84 | 75.26 | 78.84 | 78.29 | 75.75 |
| Carnivora | 64.73 | 62.87 | 61.27 | 64.07 | 61.63 | 59.68 | 64.81 | 63.37 | 60.62 |
| Perissodactyla | _58.17_ | _58.55_ | _58.92_ | _58.37_ | _59.09_ | _59.62_ | _58.13_ | _58.61_ | _59.30_ |
| Cetartiodactyla | 71.49 | 69.79 | 68.42 | 70.34 | 68.30 | 66.57 | 70.93 | 69.42 | 67.21 |
| Anthropoidea | 57.75 | 55.91 | 54.84 | 55.26 | 52.41 | 50.02 | 55.52 | 53.89 | 50.81 |
| mouse/rat | 23.09 | 22.43 | 21.89 | 21.68 | 20.59 | 18.52 | 21.74 | 21.12 | 19.70 |
| Caniformia | 52.89 | 51.39 | 50.05 | 52.25 | 50.18 | 48.51 | 53.01 | 51.77 | 49.39 |
| Artiofabula | 65.65 | 64.23 | 63.11 | 64.71 | 63.00 | 61.59 | 65.35 | 63.96 | 62.18 |
| Cetruminantia | 56.77 | 55.82 | 55.23 | 56.06 | 54.95 | 54.28 | 56.67 | 55.56 | 54.58 |
| Whippomorpha | _50.78_ | _49.98_ | _49.57_ | _50.27_ | _49.44_ | _49.14_ | _50.89_ | _49.90_ | _49.31_ |
| Cetacea | _24.09_ | _22.48_ | _21.07_ | _24.00_ | _21.46_ | _19.35_ | _25.08_ | _22.82_ | _20.19_ |

Table 4. Correlation of residuals of the nine models using Laurasiatheria calibrations to their expected median values and each other. Highlighted are the two highest fits with respect to expected residuals.

| | Expt. | R | L | I | RE | LE | IE | RT | LT |
|---|---|---|---|---|---|---|---|---|---|
| Expt. | 1.000 | | | | | | | | |
| R | 0.951 | 1.000 | | | | | | | |
| L | 0.969 | 0.990 | 1.000 | | | | | | |
| I | 0.979 | 0.978 | 0.985 | 1.000 | | | | | |
| RE | 0.976 | 0.980 | 0.987 | 0.992 | 1.000 | | | | |
| LE | 0.977 | 0.975 | 0.988 | 0.988 | 0.994 | 1.000 | | | |
| IE | <span style="color:red">0.985</span> | 0.981 | 0.988 | 0.994 | 0.986 | 0.988 | 1.000 | | |
| RT | <span style="color:red">0.980</span> | 0.978 | 0.983 | 0.993 | 0.996 | 0.992 | 0.989 | 1.000 | |
| LT | 0.977 | 0.975 | 0.986 | 0.992 | 0.997 | 0.995 | 0.988 | 0.996 | 1.000 |
| IT | 0.980 | 0.976 | 0.982 | 0.996 | 0.987 | 0.984 | 0.995 | 0.992 | 0.990 |

**3.6 Properties of divergence time estimates using both sets off fossil data**

In this section all five fossil calibrations (not including Cetacea) were used. Table 5 shows the results of this fitting using the older set of calibrations. Recall that the older HT calibration is not directly justified by fossil interpretation. The stress on the fossil calibrations is spread fairly widely, with different models showing different emphasis. Within Laurasiatheria, WH was typically being deflected further from its optimal value than HR. Indeed WH is being pulled down to ages that seem somewhat improbable. There clearly seems to be a real issue within this part of the tree as all the models are also having trouble predicting the expected age of Cetacea. They all underestimate it. This is expected if this lineage's evolutionary rate has been slowing down more rapidly than the models anticipate. This is not surprising given the large body size and long generation times of modern whales and, given the size of the fossils, this probably has been a feature of this lineage for at least the last 50 million years, becoming progressively stronger with time.

Within Supraprimates, HT is variously relatively unstressed to highly stressed, with the R models stressing the least and the I rate models the most. Something similar seems to be happening with the Rodentia calibration, which is invariably being pulled higher. The RP calibration is relatively weak and does not have much effect.

The main theme of this analysis appears to be that the fit of all models is suddenly much worse than if calibrations within only one group are used. While the R models seem relatively less stressed and the I models stressed the most, this may be an artifact of how total stress is being measured. If the R models are relatively poorly fitting, then when their unconstrained penalty is set to the degrees of freedom, this may mean that any further worsening of fit is less than what a good fitting model would encounter. This requires further diagnosis. Note that from table 6, all these models show a good correlation with the expected residuals, indeed better than previously. It seems possible that none of the measures of fit being considered here are ideal. However, there is at least one more indication that the use of both sets of calibrations is having a profound effect. This is that the estimated age of the root of placentals amongst models is now more widely distributed, being from ~96 to 118 mybp. Further, the three models with the best residuals still predict somewhat different root ages (from about 97 to 109 mybp).

Allowing each model to take up different root ages by stressing the fit by no more than ~4 units (~2 lnL units) after repotimizing all other parameters (Kitazoe et al. 2007), sees the range of potential values broaden to ~90 mybp to 103.5 and 96.3 to 131.2 mybp for the 'youngest' and 'oldest' models, respectively from Table 5.

If we fit using the younger calibrations (results not shown) the general results are similar. The penalty fit has become slightly worse in all cases than shown in table 5, while the residual fit remains high in all cases. The estimated root age remains variable, but is on average about 4 myr younger than seen in table 5.

Table 5. Fitting multiple models of the rate of evolution of the rate of evolution to the tree of placental mammals. The HR, HT, WH, RP, and Rodentia nodes used the older fossil calibrations. The fit (twice ln L difference from optimal fit) of other fossil calibrations is also shown, but these were not part of the penalty function. Model fit is relative to a single calibration normalized to 32. Ages in mybp. Underlined are nodes with well described calibrations.

| Calibration\Model | R | L | I | RE | LE | IE | RT | LT | IT |
|---|---|---|---|---|---|---|---|---|---|
| HR | 0.0 | -0.5 | -2.2 | -0.1 | 0.0 | -1.0 | 0.0 | 0.0 | -0.4 |
| HT | -0.7 | -3.5 | -8.0 | -0.9 | -4.6 | -10.4 | -0.7 | -2.5 | -5.9 |
| WH | -0.9 | -2.3 | -2.6 | -2.3 | -3.4 | -3.5 | -0.8 | -3.0 | -3.1 |
| RP | 0.0 | -0.1 | -0.2 | 0.0 | -0.1 | -0.1 | 0.0 | 0.0 | -0.1 |
| Rodentia | 0.0 | -1.9 | -5.1 | 0.0 | -3.8 | -5.9 | 0.0 | -1.5 | -3.6 |
| Cetacea | -6.4 | -8.2 | -10.5 | -6.0 | -9.0 | -13.0 | -5.1 | -6.9 | -11.2 |
| Model | R | L | I | RE | LE | IE | RT | LT | IT |
| Fit | 58.78 | 86.64 | 108.72 | 74.60 | 97.78 | 123.96 | 60.54 | 81.51 | 102.13 |
| Clade ages | R | L | I | RE | LE | IE | RT | LT | IT |
| Placentalia/Root | 113.08 | 99.28 | 96.58 | 109.47 | 100.41 | 98.87 | 118.33 | 96.27 | 95.80 |
| Boreotheria | 103.53 | 91.86 | 90.24 | 99.63 | 92.75 | 91.68 | 108.38 | 89.29 | 88.95 |
| Atlantogenata | 108.32 | 95.18 | 92.67 | 104.34 | 96.14 | 94.82 | 112.58 | 92.34 | 91.89 |
| Laurasiatheria | 92.70 | 84.05 | 85.00 | 87.17 | 82.72 | 82.53 | 94.04 | 79.94 | 79.94 |
| Supraprimates | 74.37 | 78.98 | 80.46 | 82.14 | 82.59 | 82.61 | 74.20 | 79.70 | 80.22 |
| Afrotheria | 87.44 | 76.45 | 74.15 | 81.26 | 75.55 | 74.86 | 86.65 | 73.59 | 72.65 |
| Scrotifera | 87.52 | 80.28 | 81.88 | 82.12 | 78.42 | 78.36 | 88.16 | 75.96 | 76.01 |
| Eulipotyphla | 81.70 | 73.88 | 74.73 | 73.96 | 71.05 | 71.24 | 79.73 | 69.43 | 69.27 |
| Euarchonta | 72.29 | 76.83 | 78.47 | 79.09 | 80.28 | 80.53 | 72.37 | 77.56 | 78.24 |
| Glires | 67.19 | 70.65 | 73.06 | 70.86 | 73.80 | 74.72 | 67.46 | 71.38 | 72.59 |
| Afroinsectiphilla | 81.77 | 71.13 | 68.62 | 75.00 | 69.64 | 68.90 | 80.20 | 68.34 | 67.16 |
| Fereuungulata | 83.81 | 77.49 | 79.16 | 78.67 | 75.41 | 75.32 | 84.12 | 73.21 | 73.23 |
| Chiroptera | 69.51 | 64.54 | 65.73 | 65.32 | 63.32 | 62.81 | 68.94 | 62.11 | 61.63 |
| Primates | 63.13 | 65.11 | 66.80 | 64.17 | 66.22 | 67.86 | 63.21 | 64.94 | 66.35 |
| Lagomorpha | <u>43.82</u> | <u>45.50</u> | <u>47.71</u> | <u>42.47</u> | <u>44.91</u> | <u>46.72</u> | <u>43.79</u> | <u>44.06</u> | <u>45.33</u> |
| Rodentia | <u>60.00</u> | <u>61.45</u> | <u>63.84</u> | <u>60.00</u> | <u>62.84</u> | <u>64.45</u> | <u>60.00</u> | <u>61.13</u> | <u>62.69</u> |
| Ferae | 78.29 | 72.74 | 74.26 | 73.85 | 71.11 | 70.90 | 78.53 | 69.32 | 69.23 |
| Euungulata | 80.12 | 74.62 | 75.99 | 75.40 | 72.53 | 72.36 | 80.29 | 70.63 | 70.59 |
| megabats | 31.87 | 28.99 | 28.71 | 31.28 | 29.03 | 27.14 | 33.67 | 29.44 | 27.44 |
| human/tarsier | <u>61.32</u> | <u>62.95</u> | <u>64.48</u> | <u>61.54</u> | <u>63.37</u> | <u>65.10</u> | <u>61.30</u> | <u>62.51</u> | <u>63.84</u> |
| hystrich./murid_ | 52.69 | 53.29 | 55.10 | 49.69 | 51.08 | 52.26 | 51.38 | 50.80 | 51.61 |
| Carnivora | 56.66 | 53.32 | 54.21 | 55.35 | 53.73 | 53.03 | 57.76 | 52.99 | 52.25 |
| Perissodactyla | <u>58.00</u> | <u>57.14</u> | <u>56.19</u> | <u>58.37</u> | <u>57.87</u> | <u>56.75</u> | <u>58.11</u> | <u>58.13</u> | <u>57.23</u> |
| Cetartiodactyla | 66.42 | 63.61 | 64.18 | 63.57 | 62.19 | 62.09 | 66.14 | 61.11 | 61.17 |
| Anthropoidea | 36.99 | 36.80 | 36.52 | 34.19 | 32.60 | 30.80 | 35.62 | 33.65 | 32.03 |
| mouse/rat | 14.84 | 14.82 | 14.84 | 13.48 | 13.37 | 11.21 | 14.08 | 13.42 | 12.81 |
| Caniformia | 46.26 | 43.70 | 44.40 | 45.40 | 44.20 | 43.48 | 46.99 | 43.81 | 42.98 |
| Artiofabula | 61.87 | 59.74 | 60.12 | 59.66 | 58.63 | 58.55 | 61.68 | 57.90 | 57.96 |

| | | | | | | | | |
|---|---|---|---|---|---|---|---|---|
| Cetruminantia | 54.73 | 53.63 | 53.85 | 53.43 | 52.94 | 53.01 | 54.59 | 52.65 | 52.76 |
| Whippomorpha | <u>49.62</u> | <u>48.96</u> | <u>48.90</u> | <u>48.98</u> | <u>48.72</u> | <u>48.69</u> | <u>49.68</u> | <u>48.79</u> | <u>48.78</u> |
| Cetacea | <u>23.88</u> | <u>22.54</u> | <u>21.04</u> | <u>24.17</u> | <u>21.98</u> | <u>19.56</u> | <u>24.94</u> | <u>23.47</u> | <u>20.59</u> |

Table 6. Correlation of residuals of the nine models, using Laurasiatheria calibrations, to their expected median values and to each other. Highlighted are the two highest fits with respect to expected residuals.

| | Expt. | R | L | I | RE | LE | IE | RT | LT |
|---|---|---|---|---|---|---|---|---|---|
| Expt. | 1 | | | | | | | | |
| R | 0.984 | 1.000 | | | | | | | |
| L | 0.990 | 0.988 | 1.000 | | | | | | |
| I | 0.995 | 0.983 | 0.988 | 1.000 | | | | | |
| RE | <span style="color:red">0.997</span> | 0.983 | 0.992 | 0.992 | 1.000 | | | | |
| LE | <span style="color:red">0.996</span> | 0.974 | 0.988 | 0.992 | 0.996 | 1.000 | | | |
| IE | 0.994 | 0.980 | 0.982 | 0.995 | 0.987 | 0.987 | 1.000 | | |
| RT | 0.988 | 0.986 | 0.975 | 0.987 | 0.982 | 0.978 | 0.991 | 1.000 | |
| LT | 0.994 | 0.972 | 0.985 | 0.988 | 0.996 | 0.998 | 0.983 | 0.977 | 1.000 |
| IT | 0.995 | 0.982 | 0.984 | 0.994 | 0.991 | 0.989 | 0.997 | 0.991 | 0.987 |

## 4 Discussion

The results reinforce the view that something very odd is occurring between the fossil calibrations in Supraprimates versus Laurasiatheria. It may be tempting to suggest that the problem lies within Supraprimates, and that there are many old fossils for that group that are unsampled or unrecognized. This, however, runs up against the absolute lack of rodent teeth older than ~58 mybp. Given the stratigraphy of sites in Asia and North America, the discovery of lots of probable ancestors in slightly older strata, an ecology that suggests early rodents were generalists that quickly turned up at many sites, and both the relative indestructibility and easy identifiability of rodent teeth, all suggests this is a very good calibration.

On the other hand, the best fossil calibrations within Laurasiatheria (and Afrotheria) both involve lineages that have evolved to be large bodied and have long generation times. Indeed, whales seem to have slowed down more than any of the models allow for, and their calibrations show conflict even with the Horse/Rhino calibration. It is unfortunate that the fossil record of the other Laurasian lineages, particularly smaller bodied bats and eulipotyphlans, are so unclear. While there are older fossils that seem associated with these lineages, confirming that they are within the crown group remains problematic. Of course, this may not be a coincidence, if the ages within Laurasiatheria are indeed relatively young. Following discussions with Jeff Thorne, in the future, it may be particularly desirable to model the longevity of ancestors based on living descendants, while body size, which is significantly positively correlated with longevity, can be inferred from both descendants and fossils. It would seem that this conflict within placentals may be particularly amenable to such an approach.

The newly derived calibrations are probably an improvement over what went before.

However, the differences in the end result are not huge. One area where they are very desirable is that they much easier to numerically optimize since there are no boundaries to get stuck on. However, when using the exponential for the rodents, caution was needed, as this effectively has a boundary on its lower side and this was the direction the optimization was going. If it was not reset to a higher values between runs, it seemed to cause poorly fitting solutions to be found.

Modeling continuous distributions, particularly highly skewed ones with long tails (such as the log normal can become when the variance is large compared to the mean) can make it hard to fit all the properties desired, including mode, upper and lower confidence points and the translation which acts as a lower bound. They also demand skills that are not readily available to all paleontologists. Another limitation is that they become even more difficult to handle if a distribution is bimodal. For this reason, use of smoothed histograms are encouraged. The smoothing may be as simple as linear interpolation between midpoints of the categories used. Out towards the upper and lower tails, appropriate exponential tails might be useful (obviously the lower tail would need to be reflected, translated, and lamda chosen to drop away quickly if there was a strong lower bound). The gradient changes could be an issue for some optimizers, while a bimodal character would require numerical routines that could negotiate local optima. One instance that seems to require a somewhat bimodal distribution, is the orangutan lineage calibration (Waddell and Penny 1996, Begun 20003). While the fossils all tend to be pointing towards a maximally probably date of 15-16 mybp, the biogeographic timing of movements from Africa to Asia suggest an older time of 18 to 20 mybp.

Finally, given that the residuals do not always seem to behave as expected, it would be desirable to better understand their behavior. This suggests simulations. It might also be desirable to directly optimize the variance of the various Brownian and random walk models and calculate their full relative likelihoods. An alternative is to consider the comparison of these models using the same logic as Penalized Likelihood (Sanderson 2002). Another direction for future effort is to consider more heterogeneous models. It could be a mixture of types of Brownian motion, or it could be something like a Levy process, which is a combination of a Brownian motion with sudden jumps pulled from a secondary distribution. Such models have had some success in modeling the erratic behavior of stock markets. These appear log Brownian for runs of weeks or more, but then jump (or more aptly, fall) far more than expected.

## Acknowledgements

This work was supported by NIH grant 5R01LM008626 to PJW. Thanks to Hiro Kishino, Mary Dawson, Chris Beard, Hans Thewissen, Steve Fenner, Jeff Thorne and Andrew Lawson for helpful comments and discussions.

## Author contributions

PJW originated the research, developed methods, gathered data, ran analyses, prepared figures and wrote manuscript.

# Appendix 1

The weighted tree used as the basis of divergence time estimates. The rooting point of the placental mammals is estimated by where two marsupial outgroups (Kangaroo and Opossum) join the tree. In this case it is between the two clades Atlantogenata and Boreotheria. For the program The_Times, the tree should be edited to show a single outgroup taxon which is then removed. Thus, it is the rooted weighted subtree of placentals that is the data used here in the analyses of divergence time.

((((((Pangolin: 0.134531, (Cat: 0.061232, (Dog: 0.058411, Seal: 0.056380): 0.012767): 0.026442): 0.008448, ((Horse: 0.049600, Rhino: 0.050957): 0.022978, (Lama: 0.083456, (Pig: 0.085791, (Cow: 0.083980, (Hippo: 0.066774, (Baleen_Whale: 0.027207, Sperm_Whale: 0.053725): 0.046302): 0.009415): 0.013738): 0.008344): 0.025481): 0.005936): 0.006345, (Jamacian_fruit_bat: 0.133417, (Megabat: 0.044869, Flying_fox: 0.030986): 0.049063): 0.029769): 0.009436, (Hedgehog: 0.277914, Shrew/Mole: 0.106873): 0.024773): 0.021029, ((Tree_shrew: 0.153912, (Galago: 0.147115, (Tarsier: 0.119529, (Human: 0.093707, New_World_monkey: 0.140561): 0.072800): 0.004058): 0.017025): 0.003023, ((Pika: 0.104635, Rabbit: 0.090144): 0.048364, (Squirrel: 0.113329, (Guinea_pig: 0.172194, (Mouse: 0.060919, Rat: 0.061470): 0.152141): 0.023661): 0.016468): 0.011905): 0.013773): 0.012697, (Armadillo: 0.154321, (Elephant: 0.158307, (Aardvark: 0.095692, Tenrec: 0.226400): 0.010837): 0.036863): 0.007277): 0.294538, (Kangaroo: 0.097687);